# Mixed Magnetism for Refrigeration and Energy Conversion

N.H. Dung[1], Z.Q. Ou[1], L. Caron[1], L. Zhang[1], D.T. Cam Thanh[1], G.A. de Wijs[2], R.A. de Groot[2], K.H.J. Buschow[1], E. Brück[1]*

[1] *Fundamental Aspects of Materials and Energy, Faculty of Applied Sciences, Delft University of Technology, 2629 JB Delft, The Netherlands*

[2] *Electronic Structure of Materials, Faculty of Science, Radboud University, 6525 AJ Nijmegen, The Netherlands*

* To whom correspondence should be addressed. E-mail: e.h.bruck@tudelft.nl

**The efficient coupling between lattice degrees of freedom and spin degrees of freedom in magnetic materials can be used for refrigeration and energy conversion. This coupling is enhanced in materials exhibiting the giant magnetocaloric effect. First principle electronic structure calculations on hexagonal MnFe(P, Si) reveal a new form of magnetism: the *coexistence of strong and weak magnetism in alternate atomic layers*. The weak magnetism of Fe layers (disappearance of local magnetic moments at the Curie temperature) is responsible for a strong coupling with the crystal lattice while the strong magnetism in adjacent Mn-layers ensures Curie temperatures high enough to enable operation at and above room temperature. Varying the composition on these magnetic sublattices gives a handle to tune the working temperature and to achieve a strong reduction of the undesired thermal hysteresis. In this way we design novel materials based on abundantly available elements with properties matched to the requirements of an efficient refrigeration or energy-conversion cycle.**

## 1. Introduction

Limited resources and the wish for improved prosperity call for efficient use of energy. The UN Advisory Group on Energy and Climate Change recommends a target of 40 % improved efficiency by 2030.[1] Materials research can contribute significantly to reach this target. Magnetic refrigeration offers potential to achieve a 50% higher energy-efficiency compared to vapour compression refrigeration.[2] This makes magnetic refrigeration a technology that attracts growing attention. Magnetic refrigeration is based on the magnetocaloric effect (MCE); solid magnetic materials heat up or cool down when an external magnetic field is applied or removed, due to the entropy transfer $\Delta S$ between the magnetic system and the crystal lattice. In a reversed process heat can be transformed into electricity with a thermomagnetic generator.[3] This device can be used to generate electricity from 'waste heat', heat that currently is released unused into the environment. Efficient magnetocaloric materials could therefore contribute significantly to a reduction in energy consumption. However, up to now no suitable magnetic materials were available to operate between room temperature and 400K.

Following the discovery of giant MCE in $Gd_5(Si,Ge)_4$,[4] a number of magnetocaloric materials[4-10] with a first-order magnetic phase transition (FOMT) have been intensively explored. In these materials, the FOMT enhances the magnetocaloric effect in the vicinity of the magnetic phase transition. The maximum isothermal entropy change is therefore often significantly greater than that of the benchmark material Gd (Ref. 11) that presents a second-order magnetic phase transition. Combining giant magnetocaloric materials with different $T_C$ in series, a higher efficiency and a greater temperature span than that of Gd is obtained.[12] For optimal performance the materials used in such a composite regenerator need to have very similar magnetocaloric properties, to achieve a constant entropy change as function of temperature.[13] The large thermal hysteresis frequently associated with the FOMT seriously hampers the application in a refrigeration cycle.[14] Thermal and field hystereses are intrinsic properties of a first-order material. How-ever, the size of this hysteresis may strongly depend on microstructure or strain in the system. Here we report on a novel mechanism derived from first principle electronic structure calculations. The intercalation of weak and strong magnetism in adjacent lattice planes induces a large magneto-elastic effect and giant magnetocaloric effects. With this mechanism we can generate exceptionally favourable magnetocaloric properties in broad regions of the Mn-Fe-P-Si system.

## 2. Results and Discussion

$Fe_2P$ and related alloys have attracted attention for quite some time,[15] on replacing more than 10 % of P by Si in $Fe_2P$ the hexagonal crystal lattice is transformed into an orthorhombic one.[16] To avoid the hexagonal to orthorhombic transformation we replaced Fe on the 3$g$ sites by Mn to form a compound with very interesting properties. $MnFeP_{0.5}Si_{0.5}$ with a FOMT near room temperature, crystallizes in the hexagonal $Fe_2P$ type structure that has four distinct lattice sites, the thermal hysteresis $\Delta T_{hys}$ of 35 K makes this material however unsuitable for applications.[17] In contrast to most other magnetocaloric materials the volume change in this material is rather small and we observe mainly a change in $c/a$ ratio of the hexagonal lattice. In order to elucidate the origin of the observed magnetocaloric effect, electronic structure calculations were performed on the ferromagnetic ground-state, while the behaviour at the Curie temperature was modelled by a supercell obtained by doubling the unit cell (allowing for antiferromagnetic configurations). The calculations show that layers occupied by manganese are strongly magnetic; implying that the magnetic order only is lost at the Curie temperature. The size of the Mn moment is reduced from 2.8 $\mu_B$ in the ferromagnetic phase to 2.6 $\mu_B$ in the paramagnetic phase. By contrast, the iron-layers show weak itinerant magnetism: here the Fe moment in the ferromagnetic phase is 1.54 $\mu_B$, while in the paramagnetic phase it vanishes (~0.003 $\mu_B$). This implies that the electron density around the Fe sites changes drastically at the phase transition. This change is especially significant within the Fe layer as illustrated in Figure 1. In this figure we show the difference in electron density between the ferromagnetic and the paramagnetic state. The dominant changes occur close to the Fe sites while near the Si/P sites the electron density is hardly affected. In the ferromagnetic state high electron density forms a dumbbell pointing into the empty space between adjacent Si/P atoms, while in the paramagnetic state high electron density forms a clover four pointing towards the nearest Si/P neighbours. This redistribution of electron density means that non-bonding electron density at the Fe site below $T_C$ changes into a distribution which is hybridized with Si/P above $T_C$. This change in hybridization causes the distinct change in c/a ratio observed experimentally in the magneto-elastic transition at $T_C$.

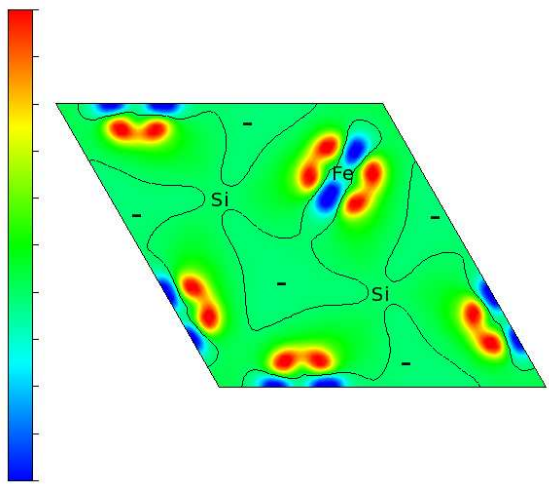

Figure 1. Difference between the electron densities calculated for situations above and below the Curie temperature in the iron-silicon/phosphorous plane (ferromagnetic density (-) substracted from the paramagnetic one). This results locally in negative electron densities where the highest ferromagnetic density was. The colour codes for the absolute electron density ranges from -4.55 × 10$^{-5}$ (dark blue) to 9.1 × 10$^{-5}$ electrons/Å$^3$ at the red end of the scale. The dark lines indicate no change in electron densities.

The loss of moments on the iron site is also clear from the partial density of states as function of energy shown in Figure 2. It shows identical curves for the two spin directions for iron above the Curie temperature, in sharp contrast with manganese that maintains its moment.

Such a combination of strong and weak magnetism in *one and the same* compound is unexpected. Weak magnetism is rare; it is found in materials like $ZrZn_2$ (Ref. 18) or $Ni_3Al$.[19] Curie temperatures are low (for example, $ZrZn_2$ :33 K; $Ni_3Al$: 23-58 K depending on composition). To the best of our knowledge no other cases of mixed magnetism have been described before. It is directly related to the giant magneto-caloric effect, because in solids the existence of magnetic moments competes with chemical bonding.[20] The loss of the magnetic moments of iron enables the strong coupling to the lattice above the Curie temperature resulting in the discontinuity of the *c/a* ratio leading to the FOMT. On the other hand, the strong magnetism of the manganese layers ensures a Curie temperature near room temperature.

Experimentally we studied the effect of changing the lattice sites occupations. The hexagonal $Fe_2P$ type of structure is found to be stable over a broad range of compositions. Similar to the orthorhombic compound[16] $T_C$ increases with increasing Si content. This increase in $T_C$ is probably caused by an increase in magnetic moments that is seen from our theoretical calculations. These calculations indicate that mainly the Fe moment on 3$f$ sites is enhanced. Neutron diffraction experiments are planned to test this finding experimentally. When increasing the Mn content in MnFe(P,Si) compounds, $T_C$ decreases (see Figure 3a). This finding may be astonishing at the first glance as Mn appears to carry a larger moment than Fe. However, the excess Mn atoms have to occupy the 3$f$ sites now where they can develop only a relatively low moment of 1.25 $\mu_B$ compared with 1.54 $\mu_B$ of Fe atoms on these sites. Additionally, it is well known that Mn alloys tend to order antiferromagnetically below a critical Mn – Mn distance. The distance between 3$g$ and 3$f$ sites is clearly below this critical distance and therefore Mn on 3$f$ sites will not contribute to the strong ferromagnetism. On the other hand an increase of the Fe content results in an increase of $T_C$. The excess Fe will occupy the 3$g$ sites where Fe always carries a large magnetic moment, and $T_C$ of hexagonal $Fe_2P_{0.8}Si_{0.2}$ 510 K exceeds the $T_C$ of the Mn containing alloy.[16]

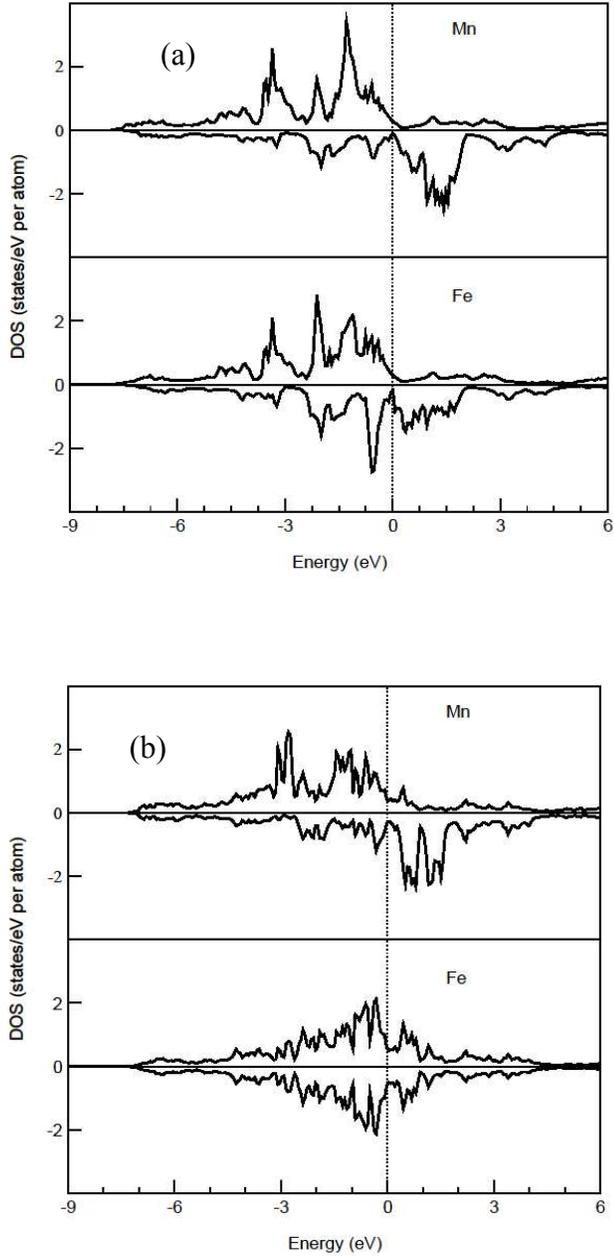

Figure 2. Partial local densities of states for the magnetic atoms in the ferromagnetic state (a) and the paramagnetic state (b) representing the state above the Curie point. Note the identical densities of state for the iron above the Curie temperature.

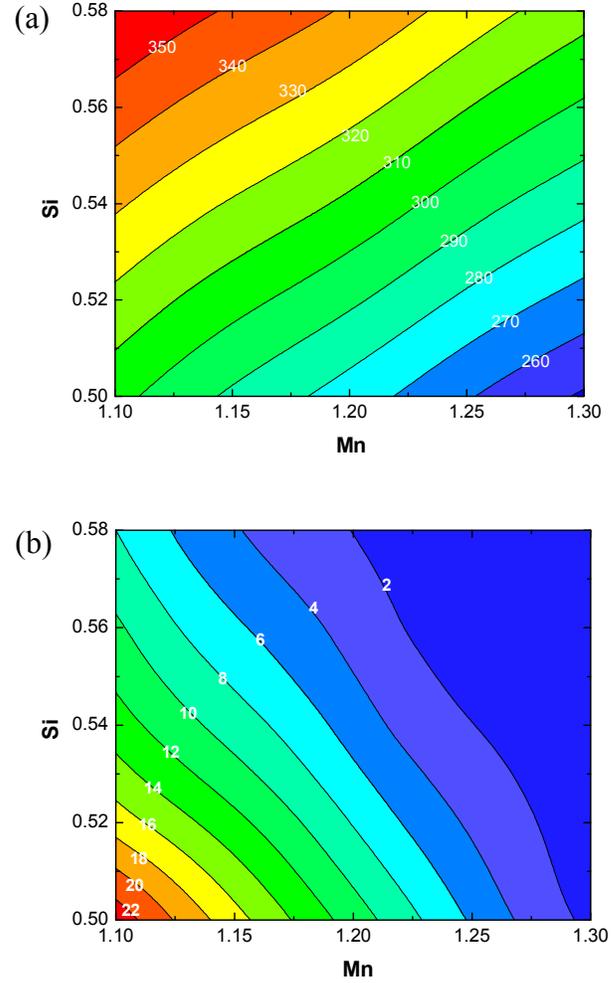

Figure 3. Partial phase diagram of the quaternary (MnFePSi) system (a) illustrating the composition dependence of the magnetic ordering temperature $T_C$ (K) for $Mn_xFe_{2-x}P_{1-y}Si_y$ compounds. (b) Composition dependence of the thermal hysteresis $\Delta T_{hys}$ (K) for $Mn_xFe_{2-x}P_{1-y}Si_y$ compounds.

Both Fe substitutions on the Mn sublattice or Mn substitutions on the Fe sublattice, as well as an increase in Si content, are beneficial in that they give rise to a decrease in $\Delta T_{hys}$ (Figure 3b). From these trends we derive that a large $\Delta S_m$ coupled with a small $\Delta T_{hys}$ can be obtained by balancing the Mn:Fe ratio and the P:Si ratio. Furthermore, $T_C$ can be tuned by changing the Mn:Fe and P:Si ratios simultaneously to keep both a large $\Delta S_m$ and a small $\Delta T_{hys}$. These trends also hold for slightly non-stoichiometric compounds. By concurrently changing Mn:Fe and P:Si ratios in $Mn_xFe_{1.95-x}P_{1-y}Si_y$ compounds, the working temperature can be controlled between 210 and 430 K for $x$=1.35, $y$ = 0.46 and $x$ = 0.66 and $y$ = 0.42, respectively, while the transition remains steep and the $\Delta T_{hys}$ remains small (1-1.5 K).

The entropy changes as function of temperature, derived from magnetic isotherms through the Maxwell relations,[5] are displayed in Figure 4. The absolute value of $\Delta S_m$ reaches 18 Jkg$^{-1}$K$^{-1}$ around both 215 and 350 K, under a magnetic field change of 0-2 T. The peak values are rather stable (between 12.8-18.3 Jkg$^{-1}$K$^{-1}$) throughout the whole temperature range from 220 to 380 K. These values are about 4 times greater than that of Gd (see the data included in Figure 4) for tuneable temperatures. Note that for the same effect more than twice the field change namely 0-5 T were required for MnFe(P,As) reported earlier.[7] Thus, with the current materials much cheaper magnets may be employed in magnetocaloric refrigerators. Because the large effect is observed over a broad range of compositions, one can achieve an equally large MCE over a wide temperature interval by cascading several alloys with slightly different compositions in one active magnetic regenerator.[13] Another important parameter to characterize the magnetocaloric effect is the adiabatic temperature change. From specific heat measurements in applied magnetic field we derive for $Mn_{1.24}Fe_{0.71}P_{0.46}Si_{0.54}$ an adiabatic $\Delta T$ of about 3 K at 320 K for a field change of 1T. This result is very close to earlier results on $Mn_{1.1}Fe_{0.9}P_{0.47}As_{0.53}$ and Gd.[4,5]

Because the effect is still large above the boiling point of water, the materials can be used in magnetocaloric generators (Figure 5) based on the Faraday induction law to transform abundant waste heat into electricity. This generator[21] shall also contain a series of different materials as depicted in Figure 4 to utilize the full temperature span from room temperature up to the temperature of the heat source. As can be seen from Figure 4, to achieve a really constant magnetic entropy change, one will need to prepare many different materials with only slightly varying composition. Because the entropy change observed in a field change from 0-1 T is already rather large, this generator without moving parts can work efficiently. This is due to a magnetic-field-induced transition with a very small magnetic hysteresis, which occurs in these compounds at low fields, as displayed in Figure 6 for the compound $Mn_{1.30}Fe_{0.65}P_{0.50}Si_{0.50}$. The extremely small magnetic hysteresis is in line with the observed small $\Delta T_{hys}$, indicating a low energy barrier for nucleation of the FOMT.

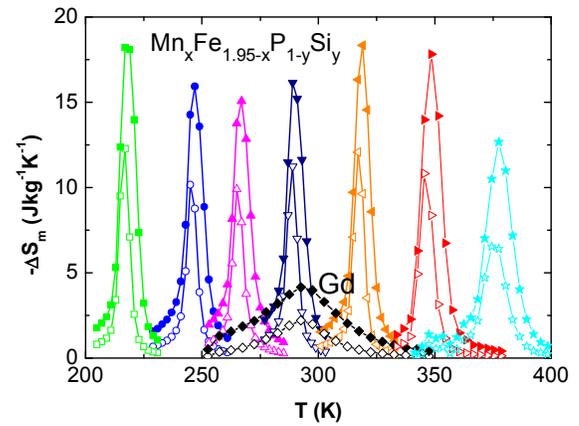

Figure 4. Isothermal magnetic entropy change under a field change of 0-1 T (open curves) and 0-2 T (solid curves) for some typical $Mn_xFe_{1.95-x}P_{1-y}Si_y$ compounds with from left to right $x$=1.34, 1.32, 1.30, 1.28, 1.24, 0.66, 0.66 and $y$ = 0.46, 0.48, 0.50, 0.52, 0.54, 0.34, 0.37, respectively. The data of Gd metal under a field change of 0-1 T (open diamond) and 0-2 T (solid diamond) are included.

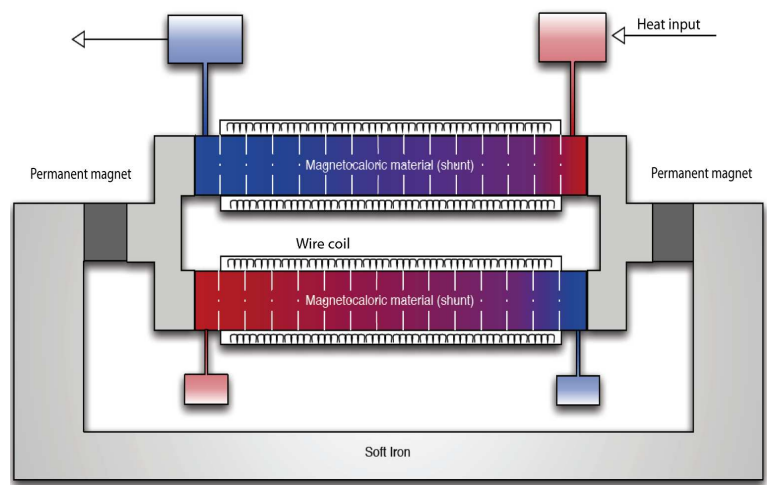

Figure 5. Concept of a magnetocaloric generator with an active shunt consisting of a series of materials with gradual changing Curie temperatures (adapted version of a design by Brioullin[21]).

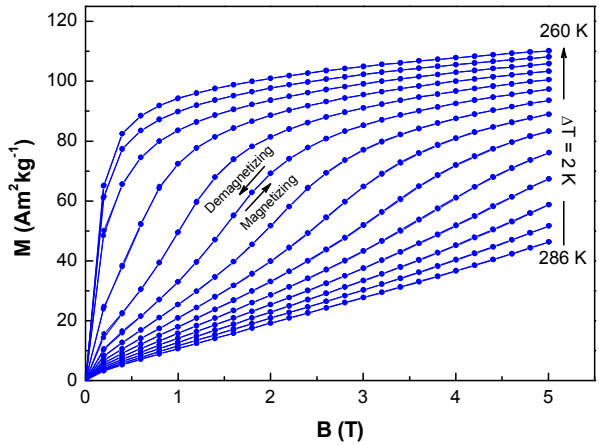

Figure 6. Magnetic isotherms of $Mn_{1.30}Fe_{0.65}P_{0.50}Si_{0.50}$ in the vicinity of the Curie temperature.

## 3. Conclusions

Combining weak and strong magnetism in a single material opens the possibility to effectively couple spin and lattice degrees of freedom. In this way we simultaneously achieved giant magnetocaloric effect and a small thermal hysteresis in Mn-Fe-P-Si compounds of hexagonal $Fe_2P$-type structure by varying Mn:Fe, P:Si. We demonstrate that the working temperature can be controlled between 210 and 430 K by concurrently changing Mn:Fe and P:Si ratios. The combination of several alloys with slightly different compositions in one active magnetic regenerator will allow for efficient magnetic refrigeration with large temperature span. The fact that we use materials that are not only globally-abundant and non-toxic but can also be industrially-mass-produced via a simple powder-metallurgical method, makes Mn-Fe-P-Si compounds particularly attractive. The discovery of these high-performance low-cost magnetic refrigerants paves the way for commercialization of magnetic refrigeration and magnetocaloric power-conversion. Additionally, the insight into the importance of the coexistence of strong and weak magnetism enables us to search specifically for novel magnetocaloric materials.

## 4. Experimental Section

First principle electronic structure calculations were performed using the localized spherical wave method (LSW)[22] using the scalar relativistic Hamiltonian. The local density exchange and correlation potential was used inside space filling and therefore overlapping spheres around the atomic constituents. Self-consistency was assumed when changes in the local partial charges were below $10^{-5}$. The calculations simulating temperatures above the Curie-point employed supercells in the *xy* and *z* directions with antiferromagnetic (*z*) or ferrimagnetic (*xy*) starting magnetic configurations. Subtle mixing prevented falling back into the ground state. The visualization of the change in the electron density between the ferromagnetic and paramagnetic state used the VESTA[23] program with VASP[24] data as input.

Mn-Fe-P-Si compounds were prepared by ball milling and solid-state reactions from starting materials consisting of Mn (99.9 %), Si (99.999 %) chips, the binary compound $Fe_2P$ (99.5 %) and red-P (99.7 %) powder. The powder obtained after milling for 10 hours was pressed into tablets. The tablets were sealed under Ar in quartz ampoules then sintered at 1373 K for 2 hours and annealed at 1123 K for 20 hours before slow cooling in the oven to room temperature. Powder X-ray diffraction patterns were obtained in a PANalytical X-pert Pro diffractometer with Cu Kα radiation, secondary flat crystal monochromator and X'celerator RTMS Detector system. Magnetic measurements were carried out using the RSO mode in a SQUID (Superconducting Quantum Interference Device) magnetometer (Quantum Design MPMS 5XL). The instrument's thermal lag of about 0.4 K with 1 K/min sweep rate around room temperature was obtained by measuring a Gd sample (3N Alfa Aesar) mounted in the same way as the other samples mentioned above. This thermal lag is not corrected for in the data displayed.

**Acknowledgements**

We thank Jack Voncken and Remco Addink for help with electron probe micro analysis and the artwork of figure 5, respectively. This work is part of an Industrial Partnership Programme IPP I18 and Materials-specific Theory for Interface and Nanoscience programme 88 of the 'Stichting voor Fundamenteel Onderzoek der Materie (FOM)' which is financially supported by the 'Nederlandse Organisatie voor Wetenschappelijk Onderzoek (NWO)'. The Industrial Partnership Programme is co-financed by BASF Future Business.